\documentclass[twocolumn,superscriptaddress,groupedaddress]{revtex4} 

\usepackage{color}
\usepackage{pstool}
\usepackage{amssymb}
\usepackage{units}
\usepackage{graphicx}
\usepackage{amsmath}
\usepackage{bm}
\usepackage{abbrevs}
\usepackage{xspace}

\usepackage[T1]{fontenc}
\usepackage[utf8]{inputenc}

\newcommand{\ket}[1]{| #1 \rangle}

\begin{document}

\title{Quantum optical signal processing in diamond}

\date{\today}

\author{Kent~A.G.~Fisher}
\affiliation{Institute for Quantum Computing and Department of Physics \& Astronomy, University of Waterloo, 200 University Avenue West, Waterloo, Ontario N2L 3G1, Canada}

\author{Duncan~G.~England}
\affiliation{National Research Council of Canada, 100 Sussex Drive, Ottawa, Ontario, K1A 0R6, Canada}

\author{Jean-Philippe~W.~MacLean}
\affiliation{Institute for Quantum Computing and Department of Physics \& Astronomy, University of Waterloo, 200 University Avenue West, Waterloo, Ontario N2L 3G1, Canada}

\author{Philip~J.~Bustard}
\affiliation{National Research Council of Canada, 100 Sussex Drive, Ottawa, Ontario, K1A 0R6, Canada}

\author{Kevin~J.~Resch}
\affiliation{Institute for Quantum Computing and Department of Physics \& Astronomy, University of Waterloo, 200 University Avenue West, Waterloo, Ontario N2L 3G1, Canada}

\author{Benjamin~J.~Sussman}
\affiliation{National Research Council of Canada, 100 Sussex Drive, Ottawa, Ontario, K1A 0R6, Canada}
\affiliation{Physics Department, University of Ottawa, 150 Louis Pasteur, Ottawa, Ontario, K1N 6N5, Canada}

\begin{abstract} 
\noindent Controlling the properties of single photons is essential for a wide array of emerging optical quantum technologies spanning quantum sensing~\cite{Giovannetti2004}, quantum computing~\cite{Knill2001}, and quantum communications~\cite{Duan2001}.  Essential components for these technologies include single photon sources~\cite{Kurtsiefer2000}, quantum memories~\cite{Kozhekin2000},  waveguides~\cite{Politi2008}, and detectors~\cite{Lita2008}. The ideal spectral operating parameters (wavelength and bandwidth) of these components are rarely similar; thus, frequency conversion and spectral control are key enabling steps for component hybridization~\cite{Kielpinski2011}. Here we perform signal processing of single photons by coherently manipulating their spectra via a modified quantum memory.  We store 723.5\,nm photons, with 4.1\,nm bandwidth, in a room-temperature diamond crystal~\cite{England2015}; upon retrieval we demonstrate centre frequency tunability over 4.2 times the input bandwidth, and bandwidth modulation between 0.5 to 1.9 times the input bandwidth. Our results demonstrate the potential for diamond, and Raman memories in general~\cite{Nunn2007,Reim2010,Bustard2013,Michelberger2015}, to be an integrated platform for photon storage and spectral conversion.
\end{abstract}

\maketitle

\noindent Spectral control is a mature field in ultrafast optics where phase- and amplitude-shaping of a THz-bandwidth pulse can be achieved using passive pulse-shaping elements in the Fourier plane~\cite{Weiner2000}. Meanwhile, a range of nonlinear optical techniques~\cite{Boyd2008} such as second harmonic generation, sum- and difference-frequency generation, four-wave mixing and Raman scattering are routinely employed to shift the frequency of laser pulses. Extending these frequency conversion techniques into the quantum regime is a critical task in the field of quantum communication but is made difficult by the low intensity of single photons, and the sensitivity of quantum states to loss and noise. Despite these challenges, quantum frequency conversion~\cite{Kumar1990a} has been demonstrated in a number of systems including waveguides in nonlinear crystals~\cite{Rakher2010, Rakher2011, Tanzilli2005, Guerreiro2014}, photonic crystal fibres~\cite{McGuinness2010}, and atomic vapour~\cite{Dudin2010}. Similarly, photon bandwidth compression has been shown using chirped pulse upconversion~\cite{Lavoie2013}.  Full control over the spectral properties of single photons~\cite{Raymer2012} has been proposed using second-~\cite{Kielpinski2011, Brecht2011} and third-order~\cite{McKinstrie2012} optical nonlinearities.

In this letter, we introduce the concept of using a Raman quantum memory to perform \emph{quantum optical signal processing}; we manipulate the properties of THz-bandwidth photons using a memory in the optical phonon modes of diamond~\cite{England2015}. In conventional signal processing~\cite{Oppenheim1989}, for example, incoming acoustic waves are converted into electronic signals by a receiver; these electronic signals are then processed before being converted back into a modified acoustic wave by a transmitter. The receiver and transmitter act as transducers, reversibly converting the information between sound waves and electronic signals.  The quantum optical analog to this device, a {\em quantum optical signal processor}, ideally manipulates the frequency and bandwidth of a single photon.  Crucially, the quantum information encoded in the photon must be maintained even while the carrier frequency and bandwidth are modified. In our demonstration, the diamond memory acts as a quantum transducer: a signal photon is mapped into an optical phonon by the write pulse, and then retrieved with its spectral properties modified according to the properties of the read pulse.

\begin{figure*}[t]

\center{\includegraphics[width=1.0\linewidth]{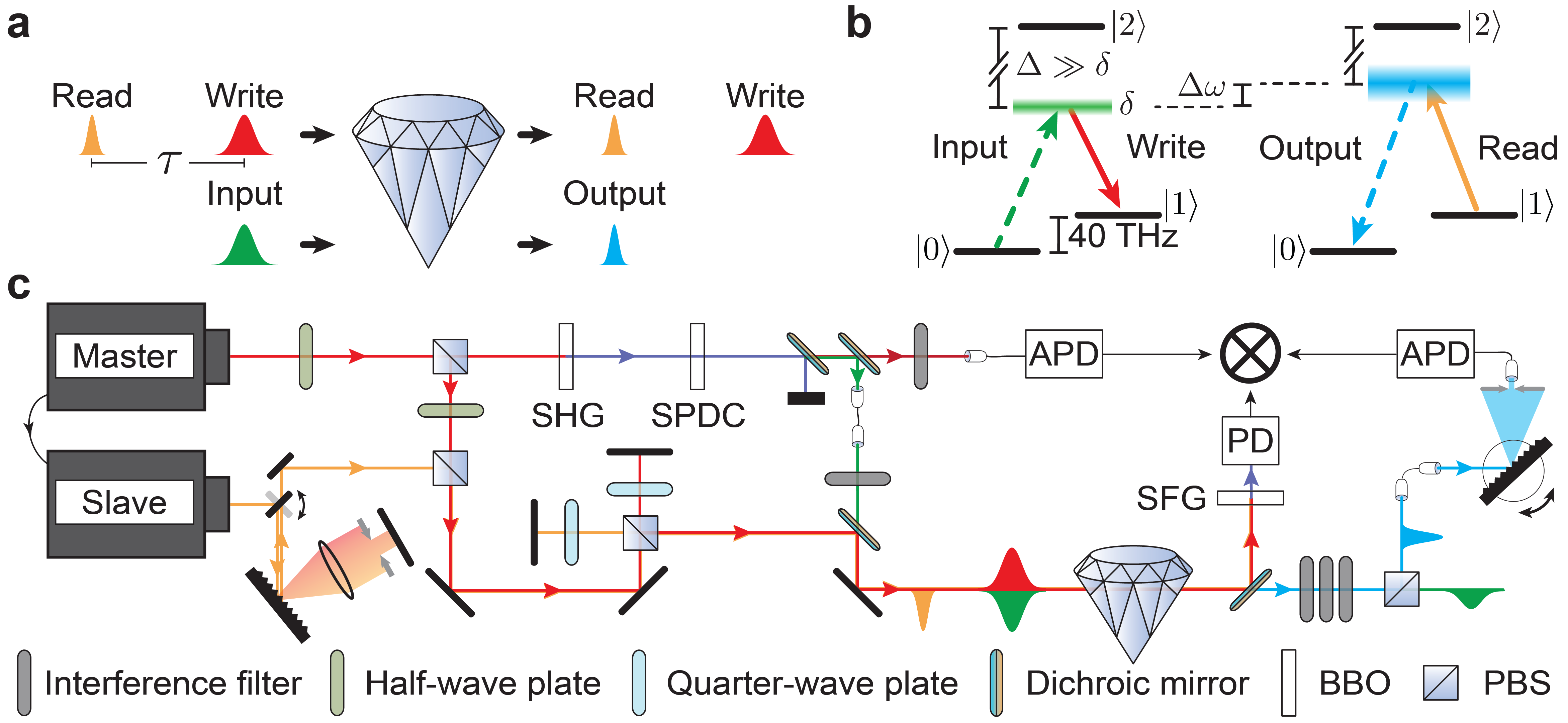}}
\caption{ {\bf Concept and experiment.} {\bf a}, Input signal photons are incident on the diamond with strong write and read fields, and are output with modified spectra. {\bf b}, Photons are Raman-absorbed to create optical phonons ($\ket{1}$), 40 THz above the ground state ($\ket{0}$). A read pulse of tunable wavelength and bandwidth retrieves the photon, determining its spectrum. Here, $\Delta$ is the detuning from the conduction band ($\ket{2}$), $\delta$ is the input photon bandwidth, $\Delta \omega$ is the detuning between input and output frequencies. {\bf c}, The master laser (red) is split between the write field and photon source, and heralded input signal photons (green) are absorbed into the diamond. The slave laser (orange) emits the read field which retrieves the photon after time $\tau$. The output signal photon (blue) spectrum is measured on a monochromator. The SFG signal from read and write pulses triggers the experiment. Coincident detections of output, herald photons and SFG events are measured by a coincidence logic unit.
 \label{fig:Setup}
 }
\end{figure*}

The quantum optical signal processor is based on a quantum memory~\cite{England2015} modelled by the $\Lambda$-level system shown in Fig.~\ref{fig:Setup}(b) where an input signal photon (723.5\,nm centre wavelength and bandwidth $\delta=4.1$\,nm full width at half maximum (FWHM)) and a strong write pulse (800\,nm, 5\,nm FWHM) are in Raman resonance with the optical phonon band (frequency 40\,THz).  The large detuning of both fields from the conduction band (detuning $\Delta \approx 950$\,THz) allows for the storage of high-bandwidth photons while the memory exhibits a quantum-level noise floor even at room temperature~\cite{England2015}. The input signal photon is transduced into the memory by Raman absorption with the write pulse, creating an optical phonon. After a delay $\tau$, the read pulse transduces the optical phonon back to a modified photon. By tuning the wavelength and bandwidth of the read pulse, we convert the wavelength of the input signal photon over a range of 17\,nm as well as performing bandwidth compression and expansion from 2.2\,nm to 7.6\,nm (FWHM), respectively. The diamond quantum memory is ideally suited to this task, offering low-noise signal processing of THz-bandwidth quantum signals at a range of visible and near-infrared wavelengths in a robust room-temperature device~\cite{England2013}.

The experimental setup is shown in Fig.~\ref{fig:Setup}c. The master laser for the experiment is a Ti:sapphire oscillator producing 44\,nJ pulses at a repetition rate of 80\,MHz and a central wavelength of 800\,nm. This beam is split in two parts: the photon-source pump and the write field. In the photon source, the second harmonic generation (SHG) of the laser light pumps collinear type-I spontaneous parametric downconversion (SPDC) in a $\beta$-barium borate (BBO) crystal, generating photons in pairs, with one at 723.5\,nm (\emph{input signal}) and the other at 894.6\,nm (\emph{herald}). The herald photon is detected on an avalanche photodiode (APD) while the input is spatially and spectrally filtered, and overlapped with the orthogonally polarized write pulse on a dichroic mirror. The input signal photon and write fields are incident on the $\left \langle 100 \right \rangle$ face of the diamond and the input is Raman-absorbed.

The photon is retrieved from the diamond using a read pulse produced by a second Ti:sapphire laser, the slave, whose repetition rate is locked to the master, but whose frequency and bandwidth can be independently modified. In this experiment we vary the read field wavelength between 784\,nm and 814\,nm, and its bandwidth between 2.1\,nm and 12.1\,nm FWHM.  In order to narrow the bandwidth of the read beam, a folded-grating $4f$-system~\cite{Weiner2000} with a narrow slit is used, while in all other configurations the $4f$ line is removed. The read beam is then overlapped with the write beam on a polarizing beamsplitter (PBS), arriving at the diamond a time $\tau$ after storage. The horizontally polarized read beam retrieves a vertically polarized \emph{output} photon from the diamond with spectral shape close to that of the read pulse, blue-shifted by the phonon frequency (40\,THz). 

The read and write pulses are separated from the signal photons after the diamond by a dichroic mirror; sum-frequency generation (SFG) of the pulses is detected on a fast-photodiode (PD) and used to confirm their successful overlap. The processed output photon is separated from the input beam by a PBS, is coupled into a single-mode fibre and directed to a monochromater. The spectrally-filtered output from the monochromater is coupled into a multi-mode fibre and detected on an APD. Coincident detections between output, herald, and read-write SFG events are measured; the experiment is triggered by the joint detection of a herald photon and an SFG signal. Refer to the \emph{Methods} for further details.

\begin{figure*}[t]
\center{\includegraphics[width=0.9\linewidth]{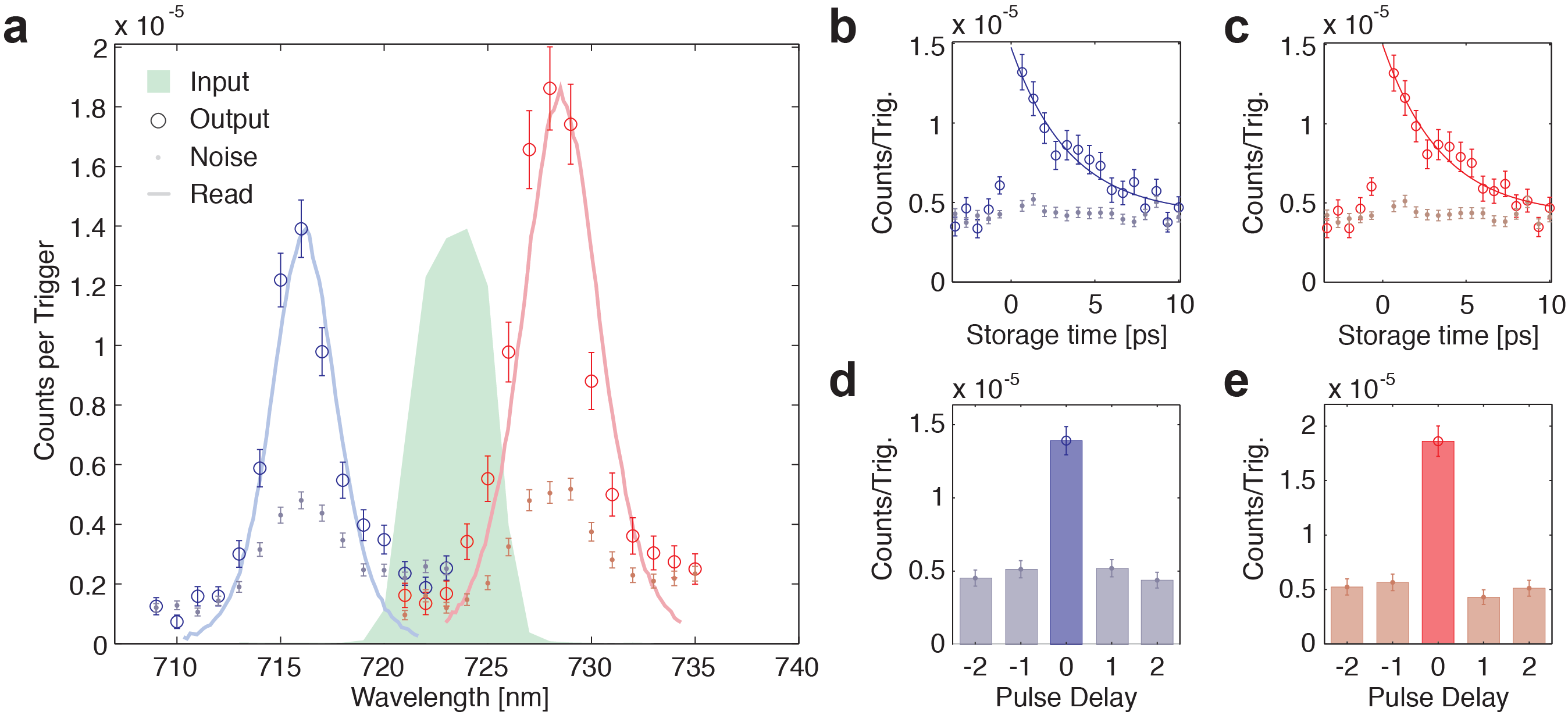}}
\caption{ {\bf Frequency conversion.} {\bf a}, The measured blue- and red-shifted output photon spectra (hollow circles), and noise (dots), when the read beam is tuned to 792\,nm and 808\,nm respectively. Corresponding read beam spectra, blue-shifted by the phonon frequency, are shown (solid lines) for reference along with the input photon spectrum (green). {\bf b}, Retrieved blue- and {\bf c}, Red-shifted signal (hollow circles), and noise (dots), as read-write delay is scanned. An exponential fit gives a phonon lifetime of 3.5\,ps. {\bf d}, Coincidence detection events between blue- and {\bf e}, red-shifted output and herald photons while scanning the electronic delay between them, as measured at the peak of the spectrum. {\bf a}--{\bf e}, Error bars show one standard deviation calculated assuming Poissonian noise.
\label{fig:FreqResults} 
}
\end{figure*}

Frequency conversion of the signal photon is observed by tuning the read laser wavelength. We vary this from 784\,nm to 812\,nm and measure the output photon wavelength using the monochromator, recording three-fold coincidence events. The resulting output spectra, with the read beam centred at 792\,nm and 808\,nm, are shown in Fig.~\ref{fig:FreqResults}a (hollow circles). The spectrum of each read pulse (solid lines), blue-shifted by the phonon frequency, is plotted along side the relevant output photon spectrum to show how the photon spectrum is determined by the read pulse. We find the peaks of the output spectra to be 716\,nm and 728\,nm with bandwidths 3.3\,nm and 3.5\,nm, FWHM respectively, making the output spectrally distinguishable from the input (green).

Following retrieval, the time-correlations characteristic of SPDC photon-pairs are preserved. This is measured by scanning the electronic delay between the signal and herald detection events in steps of 12.5\,ns (the time between adjacent oscillator pulses) and counting coincident detections. Results are shown in Fig.~\ref{fig:FreqResults}d-e for blue- and red-shifted cases respectively. We quantify this using the two-mode intensity cross-correlation function between output signal and herald fields given by $g^{(2)}_{s,h} = P_{s,h} / (P_s P_h)$. Here, $P_s (P_h)$ is the probability of detecting a photon in the signal (herald) mode, and $P_{s,h}$ is the probability of measuring a joint detection. A measurement of $g^{(2)}_{s,h} > 2$ indicates non-classical correlation~\cite{Loudon2004,Clauser1974} (see Methods), whereas uncorrelated photon detections, e.g., from noise, give $g^{(2)}_{s,h} =1$. We calculate the values of $g^{(2)}_{s,h}$ at the peak of the blue- and red-shifted spectra to be $2.7 \pm 0.2$ and $3.4 \pm 0.3$, respectively.

Figure~\ref{fig:FreqResults}b(c) shows the blue-(red-)shifted photon retrieval rate as a function of the optical delay $\tau$ between read and write pulses. An exponential function is fit to the data and we find a memory lifetime of 3.5 ps, over 12 pulse durations of the input photon (see Methods). Also plotted in Fig.~\ref{fig:FreqResults}a-c are the measured coincidences due to noise (dots), which are measured by taking the average of the $\pm12.5$ and $\pm25$~\,ns time-bins as shown in Fig.~\ref{fig:FreqResults}d and e. Noise comes from two processes: four-wave mixing~\cite{England2013}; and read pulses scattering from thermally-populated phonons producing anti-Stokes light~\cite{England2015,England2013}.  

\begin{figure} 
\center{\includegraphics[width=0.95\linewidth]{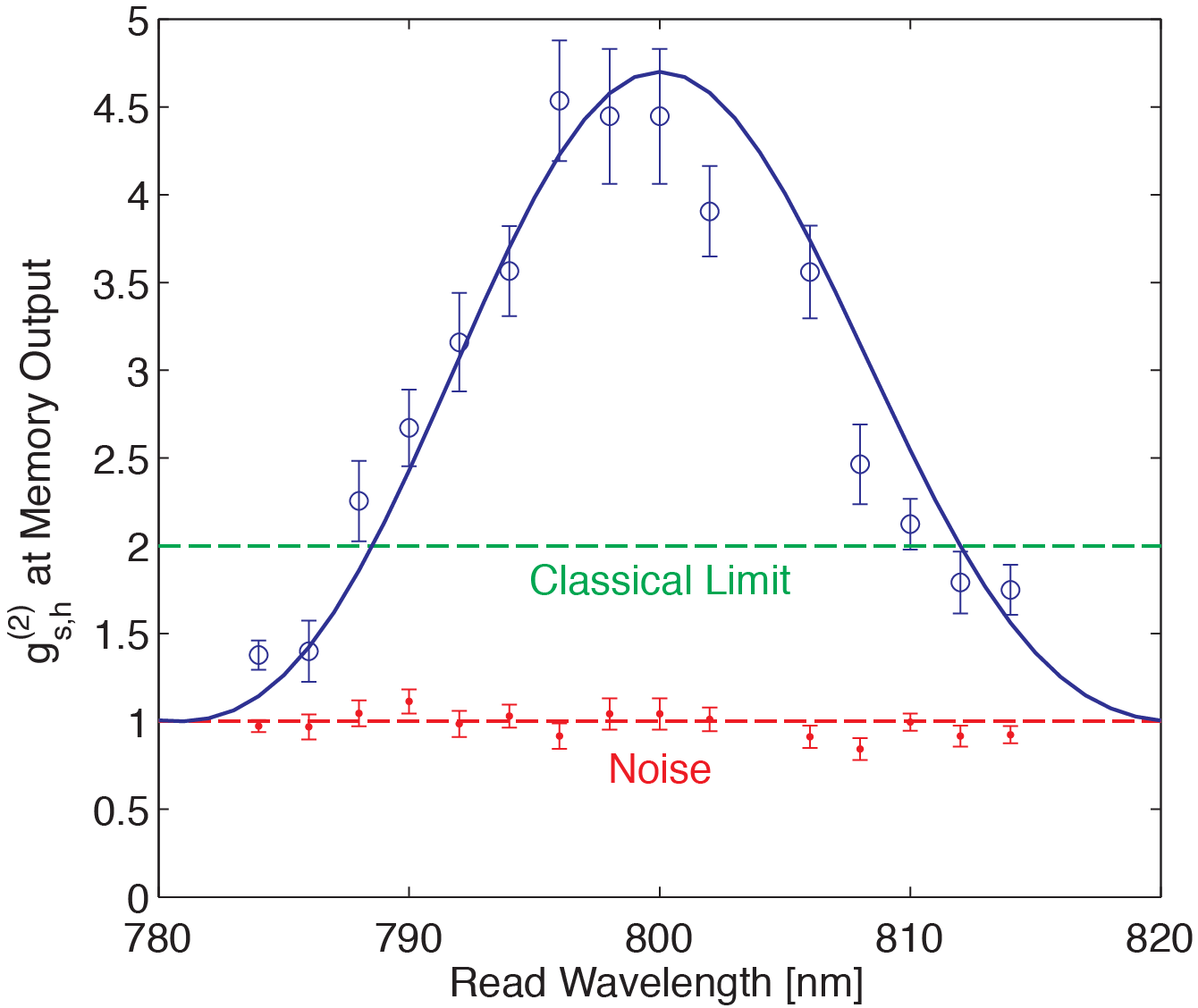}}
\caption{{\bf Range of frequency conversion.} Measured $g^{(2)}_{s,h}$ of frequency-shifted photons. Frequency conversion over an 17\,nm range is observed while maintaining non-classical statistics, i.e., $g^{(2)}_{s,h} > 2$.  The estimated $g^{(2)}_{s,h}$ (solid line)  which depends on $\text{sinc}^2(\Delta k L /2)$ agrees well with experimental data suggesting that the range of frequency conversion is determined by phase matching conditions. 
\label{fig:Sinc}
}
\end{figure}

Figure~\ref{fig:Sinc} shows $g^{(2)}_{s,h}$ measured at the peak of each output signal spectrum as the read wavelength is tuned over a 30\,nm range. We find that blue- and red-shifted single photons maintain non-classical correlations over a 17\,nm range. Since noise is uncorrelated with herald photons, we expect the noise to have cross-correlation $g^{(2)}=1$. The $g^{(2)}_{s,h}$ will increase from 1 in proportion to the signal-to-noise ratio (see Methods for further details), 

\begin{equation}
g^{(2)}_{s,h} \approx 1+\frac{ \eta_h \eta_{fc} (\Delta \omega)}{P_n}.
\label{eq:g2}
\end{equation}

\noindent Here $\eta_h = P_{s,h} / P_h = 0.13\%$ is the photon heralding efficiency in the signal arm including the monochromator, $P_n = 3.8 \times 10^{-6}$ is the probability of detecting a noise photon, and $\eta_{fc} (\Delta \omega)$ is the conversion efficiency as a function of frequency detuning, $\Delta \omega$, between input and output photons. The conversion efficiency $\eta_{fc} (\Delta \omega) = \eta_{fc} (0) \times \text{sinc}^2 (\Delta k L / 2)$ where $\eta_{fc} (0) = 1.1\%$ is the output efficiency with no frequency conversion, $L =2.3$\,mm is the length of the diamond along the propagation axis, and $\Delta k = k_i - k_o + k_r - k_w$ is the phase mismatch between the input signal$(i)$, output signal $(o)$, read $(r)$ and write $(w)$ fields due to material dispersion in diamond~\cite{England2013}. 

Inserting experimental parameters into Eq.~\ref{eq:g2} returns $g^{(2)}_{s,h} \approx 1 + 3.7 \times \text{sinc}^2 (\Delta k L / 2)$ which is plotted along side data in Fig.~\ref{fig:Sinc} (solid line). The close agreement with experiment suggests that the limitation on frequency conversion comes primarily from phase-matching conditions. We then expect that the range of frequency conversion in diamond can be extended by modifying the phase-matching conditions. We also note that the conversion efficiency can be improved with increased intensity in read and write fields, or by increasing the Raman coupling, for example in a waveguide. 

\begin{figure}[t]
\center{\includegraphics[width=0.9\linewidth]{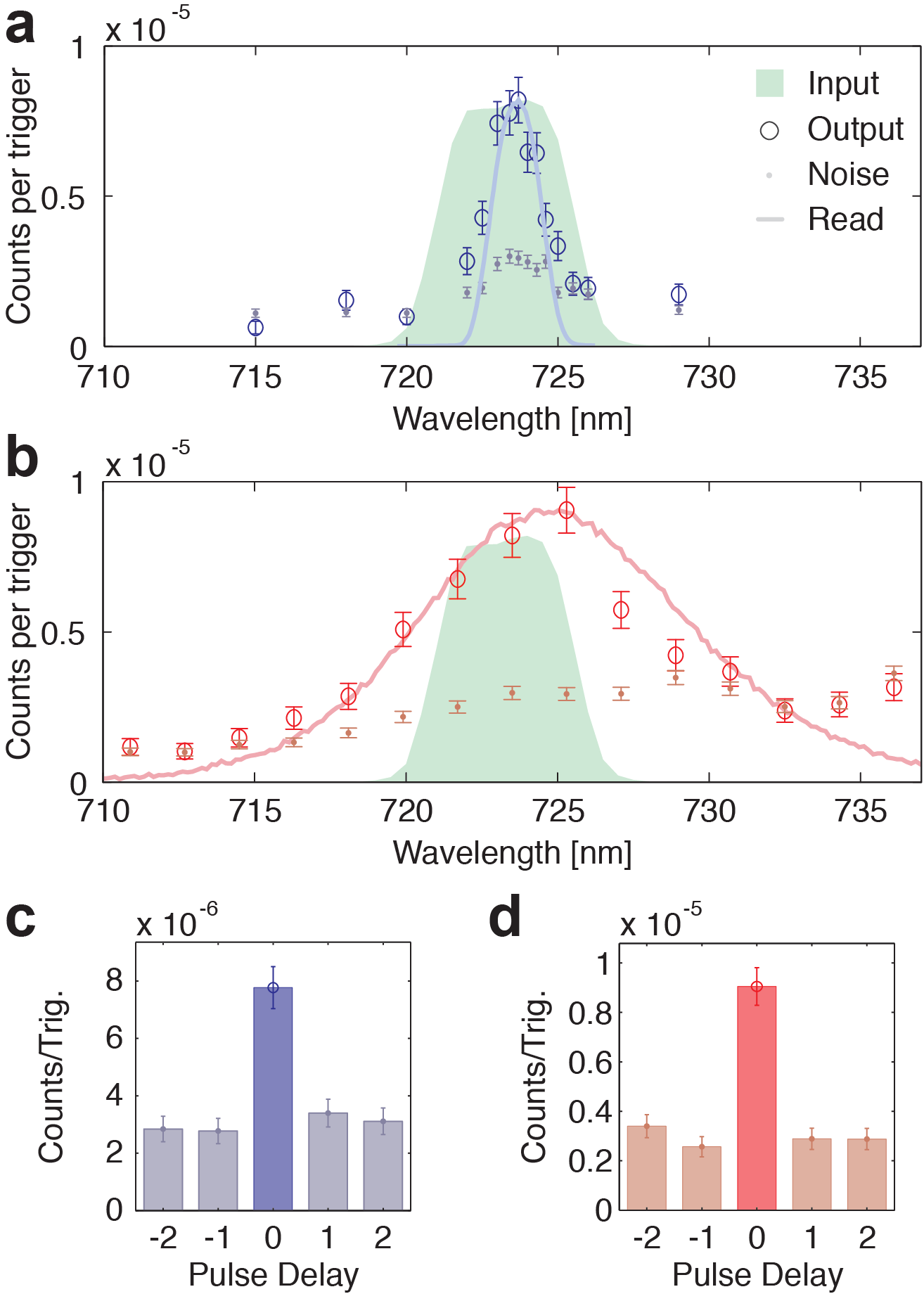}}
\caption{ {\bf Bandwidth conversion.} {\bf a}, Narrowed output spectrum (hollow circles) and noise (dots) with read beam FWHM at 2.1\,nm. {\bf b}, Expanded output photon spectrum (hollow circles) and noise (dots) with read beam FWHM at 12.1\,nm. {\bf a -- b}, Corresponding read beam spectra, blue-shifted by the phonon frequency, are shown (solid lines) for reference along with the input photon spectrum (green).  {\bf c}, Coincidence detection events between bandwidth narrowed and {\bf d}, expanded output photons and heralds while scanning the electronic delay between them, as measured at the peak of the spectrum. {\bf a}--{\bf d}, Error bars show one standard deviation calculated assuming Poissonian noise.
\label{fig:BandResults}
}
\end{figure}

Bandwidth conversion is observed by tuning the slave laser bandwidth. With the read pulse wavelength centred at 801\,nm, its bandwidth was expanded to 12.1\,nm FWHM and narrowed to 2.1\,nm FWHM using a slit in a grating $4f$ line. Figure~\ref{fig:BandResults}a(b) shows the resulting narrowed (expanded) output spectrum with the corresponding read spectrum, blue-shifted by the phonon frequency, and the input signal photon spectrum for reference. The resulting narrowed and expanded photon bandwidths are 2.2\,nm and 7.6\,nm, FWHM respectively. Figure~\ref{fig:BandResults}c(d) shows the conservation of timing correlations between bandwidth-narrowed (-expanded) photons and herald photons, respectively. We measure $g^{(2)}_{s,h} = 2.6 \pm 0.2$ in the narrowed bandwidth case, $g^{(2)}_{s,h} = 2.9 \pm 0.2$ in the expanded bandwidth case, showing that bandwidth-converted output light from the diamond maintains non-classical correlations with herald photons. 

In conclusion, we have demonstrated ultrafast quantum optical signal processing by adjusting the central wavelength, and bandwidth, of THz-bandwidth heralded single photons. Critically, the non-classical photon statistics of the single photon source are retained after signal processing. We achieve this spectral control using a modified Raman quantum memory in diamond; the single photons are stored to the memory in one spectral mode, and recalled from it in another. Diamond therefore offers low-noise THz-bandwidth storage and signal processing on a single, robust, room-temperature platform. Quantum memories for long-distance quantum communication typically demand long storage times at the expense of high-bandwidth; this application leverages a high-bandwidth memory where long-storage time is not required.

\section*{Methods}
\noindent{\bf Photon source:} Laser light is frequency-doubled by type-I SHG in a 1\,mm BBO crystal before pumping collinear type-I SPDC in a second 1\,mm BBO crystal. Horizontally-polarized photon pairs are generated at 894.6\,nm and 723.5\,nm. Remaining pump light is filtered out and photon pairs are separated by a 801\,nm long-pass dichroic mirror.  The 894.6\,nm photon passes through a 5\,nm inteference filter, is coupled into a single-mode fibre and detected on an APD. A detection heralds the presence of the 723.5\,nm photon, which is spatially-filtered in a single-mode fibre and spectrally-filtered by an interference filter with bandwidth of 5\,nm (FWHM). The input and write fields are overlapped using a 750\,nm shortpass dichroic mirror. The input and write fields are focused onto the diamond by an achromatic lens of focal length 6\,cm. 

\noindent{\bf Diamond:} The diamond is a high-purity, low birefringence crystal grown by chemical vapour deposition by Element Six Ltd. The crystal is 2.3\,mm long, cut along the $\left \langle \text{100} \right \rangle$ lattice direction and polished on two sides.

\noindent{\bf Storage time:} Absorption of the input photon by the diamond lattice is observed by an 18\% dip in input-herald coincidences when the input photon and write field arrive at the diamond simultaneously. The duration of the input photon can be deconvolved from the width of the absorption dip, 346\,fs. With write pulses 190\,fs in duration, the input photon pulse duration time is $\sqrt{ 346^2 - 190^2 } = 289$\,fs, assuming transform-limited gaussian pulses. The characteristic storage time of the diamond memory is 3.5 ps, found from an exponential fit to storage data, over 12-times the duration of the input pulse.

\noindent{\bf Laser locking:} The repetition rate of the slave laser is locked to that of the master using a Spectra Physics Lok-to-Clock device. We send read and write beams through a cross-correlator (type-II sum-frequency generation (SFG) in a 1\,mm BBO crystal) and detect the resulting signal on a fast-photodiode (PD) confirming that the time difference between the two pulses is $\leq 200$\,fs. We measure a typical SFG signal rate of 2.5\,MHz; we use this signal to trigger the experiment. 

\noindent{\bf Monochromator:} The monochromator (Acton SP2300) is comprised of a 1200\,g/mm grating between two 30\,cm focal length spherical mirrors. The output is coupled to a multi-mode fibre (105\,$\mu$m core). The apparatus has a spectral resolution of 1.1\,nm and an overall efficiency of 10\% at 723\,nm.

\noindent{\bf Cross-correlation function:} The cross-correlation function between the herald and frequency-converted light is given by $g^{(2)}_{s,h} = P_{s,h} / (P_s P_h)$. Classically, $g^{(2)}_{s,h}$ is upper-bounded by a Cauchy-Schwarz inequality~\cite{Loudon2004,Clauser1974} $g^{(2)}_{s,h} \leq \sqrt{g^{(2)}_{s,s} g^{(2)}_{h,h}}$. Here, the terms on the right-hand side are the intensity auto-correlation functions for the output signal and herald fields, which we assume, being produced by spontaneous parametric downconversion, follow thermal statistics and have $g^{(2)}_{s,s} = g^{(2)}_{h,h} =2$. Adding any uncorrelated noise would strictly lower terms on the right-hand side toward 1. To model the effect of noise on this measurement, we assume that the signal is made up of a mixture of noise photons (detected with probability $P_n = 3.8 \times 10^{-6}$) and frequency-converted photons (probability $P_\gamma$), such that

\begin{align}
P_s & = P_\gamma + P_n \approx P_h \eta_h \eta_{fc} (\Delta \omega) + P_n, \\
P_{s,h} & = P_{\gamma,h} + P_{n,h}  \approx P_h \eta_h \eta_{fc} (\Delta \omega) + P_n P_h,
\end{align}

\noindent where $\eta_{h} = 1.3 \times 10^{-3}$ is the heralding efficiency which equates to the collection efficiency of the entire signal arm, including the monochromator, and $\eta_{fc} (\Delta \omega)$ is the efficiency of the quantum frequency conversion. This returns: 

\begin{equation}
g^{(2)}_{s,h} = \frac{\eta_h \eta_{fc}(\Delta \omega) + P_n}{P_h \eta_h \eta_{fc} (\Delta \omega) + P_n},
\end{equation}

\noindent from which eq.~\ref{eq:g2} follows, given that $P_h \eta_h \eta_{fc}(\Delta \omega) \ll P_n$.

\noindent{\bf Background subtraction:} When measured at the photon source the input and herald have $g^{(2)}_{in,h} = 164$. Due to imperfect polarization extinction the input photon can, with low probability, traverse the monochromator and be detected, thereby artificially inflating the measured $g^{(2)}_{s,h}$ of the converted output. For this reason we make a measurement with no control fields present and subtract these counts from the output signal when control fields are present to portray an accurate value of $g^{(2)}_{s,h}$. 

\bibliographystyle{naturemag} 
\bibliography{freqconv}

\section*{Acknowledgments}
The authors thank Matthew Markham and Alastair Stacey of Element Six Ltd. for the diamond sample.They also thank Paul Hockett and Khabat Heshami for fruitful discussions. They thank Rune Lausten for his invaluable insights. Doug Moffatt and Denis Guay provided important technical assistance. This work was supported by the Natural Sciences and Engineering Research Council of Canada, Canada Research Chairs, the Canada Foundation for Innovation, Ontario Centres of Excellence, and the Ontario Ministry of Research and Innovation Early Researcher Award.

\section*{Author contributions}
K.A.G.F, D.G.E., and J.-P.W.M. performed the experiment and analyzed data.  All authors contributed to the final manuscript. 

\section*{Additional information}
The authors declare no competing financial interests.  Correspondence and requests for materials should be addressed to B.J.S. (ben.sussman@nrc.ca) or to K.J.R. (kresch@uwaterloo.ca).
\end{document}